\documentstyle[12pt,epsfig,graphics,cite,amssymb,amsmath,bm]{article}
\begin{document}\bibliographystyle{plain}\begin{titlepage}
\renewcommand{\thefootnote}{\fnsymbol{footnote}}\hfill
\begin{tabular}{l}HEPHY-PUB 957/16\\UWThPh-2016-2\\February
2016\end{tabular}\\[2cm]\Large\begin{center}{\bf ANALYTIC
BETHE--SALPETER DESCRIPTION OF THE LIGHTEST PSEUDOSCALAR MESONS}
\\[1cm]\large{\bf Wolfgang LUCHA\footnote[1]{\normalsize\ {\em
E-mail address\/}: wolfgang.lucha@oeaw.ac.at}}\\[.3cm]\normalsize
Institute for High Energy Physics,\\Austrian Academy of
Sciences,\\Nikolsdorfergasse 18, A-1050 Vienna,
Austria\\[1cm]\large{\bf Franz
F.~SCH\"OBERL\footnote[2]{\normalsize\ {\em E-mail address\/}:
franz.schoeberl@univie.ac.at}}\\[.3cm]\normalsize Faculty of
Physics, University of Vienna,\\Boltzmanngasse 5, A-1090 Vienna,
Austria\\[2cm]{\normalsize\bf Abstract}\end{center}\normalsize

Within the Bethe--Salpeter formalism for instantaneous
interactions, we describe, along a totally analytic route, the
lightest pseudoscalar mesons by quark--antiquark bound states
which show at least three indispensable general features, namely,
the (almost) masslessness required for pions and kaons to be
interpretable as (pseudo) Goldstone bosons, the suitable
asymptotic behaviour in the limit of large spacelike relative
momenta as determined by the relationship between quark mass
function and Bethe--Salpeter amplitudes, and a pointwise behaviour
for finite spacelike relative momenta suited for guaranteeing
colour confinement.\vspace{3ex}

\renewcommand{\thefootnote}{\arabic{footnote}}\end{titlepage}

\section{Introduction}Within particle physics, the members of the
multiplet of lightest pseudoscalar mesons,~viz., pions and kaons,
can be interpreted from two perspectives: on the one hand, they
appear~to be bound states of the fundamental degrees of freedom,
i.e., quarks and gluons, of quantum chromodynamics (QCD), the
quantum field theory which describes the strong interactions; on
the other hand, they may be regarded as the nearly massless
(pseudo) Goldstone bosons of the spontaneously (and, to a minor
degree, explicitly) broken chiral symmetries of QCD.

Quantum field theory describes bound states by means of the
covariant Bethe--Salpeter formalism \cite{BSE}. The latter's
instantaneous limit \cite{WL05:IBSEWEP}, with the Salpeter
equation \cite{SE} as its most prominent outcome, enables us to
evade obstacles arising in applications of this framework.
Recently, by use of earlier developed inversion techniques
\cite{WL13} we related, for \emph{Goldstone-type\/}
quark--antiquark bound states, Salpeter solutions compatible with
constraints arising from QCD Dyson--Schwinger equations to
\emph{configuration-space potentials\/} $V(r),$ $r=|\bm{x}|,$
encoding the impacts of strong interactions \cite{WLPoS,WL15}. In
this paper, we implement, in addition, boundary conditions imposed
by colour confinement on the solutions of the Bethe--Salpeter
equation.

The outline of this paper is as follows. In Sec.~\ref{Sec:IBSAM},
we present a rather condensed sketch of our actual route from the
Bethe--Salpeter equation to the interaction potentials responsible
for the formation of light pseudoscalar mesons acting as
pseudo-Goldstone bosons of QCD. In Sec.~\ref{Sec:DSE}, we briefly
recall how to exploit solutions of the Dyson--Schwinger
equation~for the quark propagator to derive information about the
behaviour of meson Salpeter amplitudes. In Sec.~\ref{Sec:ARP}, we
considerably refine the outcomes of Ref.~\cite{WL15} by taking
into account a particular implication of the violation of the
axiom of reflection positivity --- which acts as a sufficient but
not necessary condition for confinement --- for the propagator of
any coloured degree of freedom of QCD. In Sec.~\ref{Sec:Po}, we
discuss some of those cases that allow for analytic derivation of
the potentials. Finally, in Sec.~\ref{Sec:SCO} we collect the
insights gained by this sharpened analysis.

\section{Instantaneous Bethe--Salpeter Handling of Mesons}
\label{Sec:IBSAM}

\subsection{The Salpeter approach to fermion--antifermion bound
states}Within the framework of relativistic quantum field
theories, the bound states of elementary particles --- with total
momentum $P$ and relative momentum $p$ --- can be described by
their \emph{Bethe--Salpeter amplitude\/} $\Phi(p,P),$ governed by
the homogeneous Bethe--Salpeter equation \cite{BSE}. If the
interactions between the bound-state constituents may be
approximated by~their instantaneous limit and if the propagators
of these particles depend on the time component $p_0$ of $p$ in a
sufficiently simple form, this Bethe--Salpeter equation reduces,
upon integration over $p_0,$ to a (generic) instantaneous
Bethe--Salpeter equation \cite{WL05:IBSEWEP} for the
\emph{Salpeter amplitude\/}\begin{equation}
\phi(\bm{p})\equiv\frac{1}{2\pi}\int{\rm d}p_0\,\Phi(p)\
.\label{Eq:SA}\end{equation}Assuming, for the bound-state
constituents, free propagation with effective masses leads to the
\emph{Salpeter equation\/} \cite{SE}. Defining, for the particle
$i=1,2$ of mass $m_i$ and momentum~$\bm{p},$~by
\begin{equation}E_i(\bm{p})\equiv\sqrt{\bm{p}^2+m_i^2}\ ,\qquad
H_i(\bm{p})\equiv\gamma_0\,(\bm{\gamma}\cdot\bm{p}+m_i)\
,\qquad\Lambda_i^\pm(\bm{p})\equiv\frac{E_i(\bm{p})\pm
H_i(\bm{p})}{2\,E_i(\bm{p})}\ ,\label{Eq:EHP}\end{equation}its
free energy $E_i(\bm{p}),$ Dirac Hamiltonian $H_i(\bm{p})$ and
projection operators $\Lambda_i^\pm(\bm{p})$ for~positive or
negative energy, this Salpeter equation, \pagebreak for the bound
states of a fermion of~mass~$m_1$~and momentum $\bm{p}_1$ and an
antifermion of mass $m_2$ and momentum $\bm{p}_2,$ can be cast
into the~form\begin{align}\phi(\bm{p})&=\int\frac{{\rm
d}^3q}{(2\pi)^3} \left(\frac{\Lambda_1^+(\bm{p}_1)\,\gamma_0\,
[K(\bm{p},\bm{q})\,\phi(\bm{q})]\,\gamma_0\,\Lambda_2^-(\bm{p}_2)}
{P_0-E_1(\bm{p}_1)-E_2(\bm{p}_2)}\right.\nonumber\\[1ex]
&\hspace{11.11ex}\left.-\frac{\Lambda_1^-(\bm{p}_1)\,\gamma_0\,
[K(\bm{p},\bm{q})\,\phi(\bm{q})]\,\gamma_0\,\Lambda_2^+(\bm{p}_2)}
{P_0+E_1(\bm{p}_1)+E_2(\bm{p}_2)}\right).\label{Eq:SE}\end{align}
The integration kernel, $K(\bm{p},\bm{q}),$ subsumes the
interactions experienced by the bound-state constituents. If the
fermions couple identically, its action on $\phi(\bm{q})$ forms a
series of products of tensor products $\Gamma\otimes\Gamma$ of
matrices $\Gamma$ in Dirac space and
Lorentz-scalar~potentials~$V_\Gamma(\bm{p},\bm{q})$:
$$[K(\bm{p},\bm{q})\,\phi(\bm{q})]=\sum_\Gamma
V_\Gamma(\bm{p},\bm{q})\,\Gamma\,\phi(\bm{q})\,\Gamma\ .$$The
energy projectors in Eq.~(\ref{Eq:EHP}) entail for all solutions
(\ref{Eq:SA}) the (in fact, single \cite{WL07:HORSE}) constraint
\begin{equation}
\Lambda_1^+(\bm{p}_1)\,\phi(\bm{p})\,\Lambda_2^+(\bm{p}_2)=
\Lambda_1^-(\bm{p}_1)\,\phi(\bm{p})\,\Lambda_2^-(\bm{p}_2)=0\
.\label{Eq:EPS}\end{equation} Evidently, such three-dimensional
reduction enables us to construct a relationship between
(Poincar\'e-covariant) descriptions of bound states by means of
the Bethe--Salpeter equation and the notion of static interaction
potentials acting between the bound-state constituents.

\subsection{Light pseudoscalar mesons}The Salpeter amplitude
$\phi(\bm{p})$ of any bound state composed of a spin-$\frac{1}{2}$
fermion and a spin-$\frac{1}{2}$ antifermion with vanishing total
spin quantum number involves, upon expansion over some basis in
the ``Dirac'' space of complex $4\times4$ matrices, just two
independent components. Let us call them $\varphi_1(\bm{p})$ and
$\varphi_2(\bm{p}).$ The general form of a Salpeter amplitude
$\phi(\bm{p})$ is determined by the constraint (\ref{Eq:EPS}). It
can be read off from, e.g., Eq.~(4.9) of Ref.~\cite{JFL} or
Eq.~(12) of Ref.~\cite{OVW}; for the bound states of interest,
only three terms enter into such ``Dirac''~expansion of
$\phi(\bm{p}).$ For simplicity, we focus, in the following to the
flavour-symmetric limit of bound~states of a quark and an
antiquark of equal masses $m_1=m_2=m.$ Then, recalling the
definition (\ref{Eq:EHP}) of the Dirac Hamiltonian, two of the
three terms merge to $H(\bm{p}),$ and all $\phi(\bm{p})$
acquire~the~form\begin{equation}
\phi(\bm{p})=\left[\varphi_1(\bm{p})\,\frac{H(\bm{p})}{E(\bm{p})}
+\varphi_2(\bm{p})\right]\gamma_5\ .\label{Eq:PSA}\end{equation}

It is not an extremely daring move to assume the interaction
kernel to be of convolution type and to respect spherical
symmetry. In this case, clearly, the trivial reference to angular
variables may be separated from the dependence on the radial
momentum variables $p\equiv|\bm{p}|$ etc., the Salpeter equation
(\ref{Eq:SE}) can be reduced to an equivalent system of coupled
equations for the radial factors $\varphi_i(p)$ of the independent
Salpeter components $\varphi_i(\bm{p}),$ $i=1,2,\dots,$ and the
effective interactions may be described by central potentials
$V_\Gamma(r),$ each of which~enters the radial equations by its
Fourier--Bessel transforms \cite{JFL,OVW}. Dropping the index
$\Gamma,$ the latter read, in terms of the spherical Bessel
functions of the first kind \cite{AS} $j_i(z),$
$i=0,\pm1,\pm2,\dots,$$$V_L(p,q)\equiv8\pi\int\limits_0^\infty{\rm
d}r\,r^2\,j_L(p\,r)\,j_L(q\,r)\,V(r)\ ,\qquad p\equiv|\bm{p}|\
,\qquad q\equiv|\bm{q}|\ ,\qquad L=0,1,2,\dots\ .$$

We intend to infer (or, at least, to constrain) the potential
functions $V_\Gamma(\bm{p},\bm{q})$ from some knowledge of the
solutions $\phi(\bm{p})$ of the Salpeter equation (\ref{Eq:SE}).
In order to pose a well-defined inversion problem, the Lorentz
nature of the Dirac-matrix tensor products $\Gamma\otimes\Gamma$
entering in one's interaction kernel $K(\bm{p},\bm{q})$ must be
specified. \pagebreak As done in Refs.~\cite{WLPoS,WL15},~we adopt
for $\Gamma\otimes\Gamma$ the Fierz-symmetric linear combination
of scalar, pseudoscalar and vector~Dirac~structures
$$\Gamma\otimes\Gamma=\frac{1}{2}\,(\gamma_\mu\otimes\gamma^\mu+
\gamma_5\otimes\gamma_5-1\otimes1)\ .$$Apart from its
phenomenological importance, the advantage of this choice is the
collapse of the Salpeter equation (\ref{Eq:SE}) for a spin-singlet
bound state to two \emph{coupled\/} eigenvalue equations for the
two radial components, $\varphi_1(p)$ and $\varphi_2(p),$
determining the Salpeter amplitude (\ref{Eq:PSA})
\cite{WL07:HORSE}:\begin{align}&2\,E(p)\,\varphi_2(p)
+2\int\limits_0^\infty\frac{{\rm d}q\,q^2}{(2\pi)^2}\,V_0(p,q)
\,\varphi_2(q)=\widehat{M}\,\varphi_1(p)\ ,\nonumber\\
&2\,E(p)\,\varphi_1(p)=\widehat{M}\,\varphi_2(p)\ ,\qquad
E(p)\equiv\sqrt{p^2+m^2}\ .\label{Eq:5ess}\end{align}

For all bound states with mass eigenvalues
$\widehat{M}\equiv\sqrt{P^2}=0$ --- and thus, in particular, for
every Goldstone boson, owing to its inevitably vanishing mass ---
these equations \emph{decouple\/}: the second relation (being of
purely algebraic nature) forces $\varphi_1(p)$ to
vanish~identically, i.e., $\varphi_1(p)\equiv0,$ whereas the other
one --- an integral equation equivalent to the spinless~Salpeter
equation\footnote{For concise reviews elucidating various facets
of the spinless Salpeter equation, consult, e.g.,
Refs.~\cite{WL-SSE}.} --- governs, via $\varphi_2(p),$ the
Salpeter amplitude of any massless spin-singlet
meson:\begin{equation}E(p)\,\varphi_2(p)
+\int\limits_0^\infty\frac{{\rm d}q\,q^2}{(2\pi)^2}\,V_0(p,q)\,
\varphi_2(q)=0\ .\label{Eq:REE}\end{equation}Thus, the general
form of each spin-singlet solution $\phi(\bm{p})$ corresponding to
a vanishing mass eigenvalue $\widehat{M}$ of the Salpeter equation
(\ref{Eq:SE}) with Lorentz structure of its interaction kernel of
the Fierz-symmetric form $\Gamma\otimes\Gamma=\frac{1}{2}\,
(\gamma_\mu\otimes\gamma^\mu+\gamma_5\otimes\gamma_5-1\otimes1)$
is given by $\phi(\bm{p})=\varphi_2(\bm{p})\,\gamma_5.$ So, the
actual task is to solve the Salpeter equation (\ref{Eq:REE}) for
known potential function~$V_0(p,q)$ or, by inversion, to extract
the underlying potential $V(r)$ from~knowledge of its
solution~$\varphi_2(q).$

\subsection{Configuration-space inversion of bound-state
problem}\label{Sec:IP}Introducing, in terms of the spherical
Bessel function of the first kind $j_0(z)=(\sin z)/z$ \cite{AS},
the Fourier--Bessel transforms of Salpeter component
$\varphi_2(p)$ and kinetic term $E(p)\,\varphi_2(p)$~by
\begin{align*}&\varphi(r)\equiv\sqrt\frac{2}{\pi}
\int\limits_0^\infty{\rm d}p\,p^2\,j_0(p\,r)\,\varphi_2(p)\
,\qquad T(r)\equiv\sqrt\frac{2}{\pi}\int\limits_0^\infty{\rm
d}p\,p^2\,j_0(p\,r)\,E(p)\,\varphi_2(p)\ ,\end{align*}out of the
system of coupled relations forming the Bethe--Salpeter
quintessence (\ref{Eq:5ess}) the only member that ``survives'' the
Goldstone limit $\widehat{M}\to0,$ Eq.~(\ref{Eq:REE}), reads, in
configuration~space,$$T(r)+V(r)\,\varphi(r)=0\ .$$From such
bound-state equation, the potential $V(r)$ may be read off by
division by $\varphi(r)$ \cite{WL15}:\begin{equation}
V(r)=-\frac{T(r)}{\varphi(r)}\ .\label{Eq:Po}\end{equation}Of
course, some caution must be exercised if the Salpeter amplitude
in configuration~space, $\varphi(r),$ exhibits one or more zeros,
since, in general, each such zero will induce a singularity of the
potential $V(r)$: for instance, if $\varphi(r)$ proves to have a
single zero at $r=r_0>0$~one~might be well advised to first
consider the domain $(0,r_0)\cup(r_0,\infty)$ and then take the
limits~$r\to r_0.$ It goes without saying that such due care is
implicitly understood in the following analyses.

\section{Quark Propagator Constrains Salpeter Amplitudes}
\label{Sec:DSE} In the present context, both foundation and
primary source of information for constraining the behaviour of
the Salpeter amplitude (\ref{Eq:SA}) as a function of the relative
momentum $\bm{p}$ is the observation \cite{PM97a,PM97b} that, in
the chiral limit, the renormalized axial-vector Ward--Takahashi
identity of QCD relates the solution of the \emph{Bethe--Salpeter
equation\/} for a flavour-nonsinglet pseudoscalar meson to the
solution of the Dyson--Schwinger equation for the dressed quark
propagator $S(p),$ defined by two (real) Lorentz-scalar functions
which can be interpreted as the quark mass function, $M(p^2),$ and
quark wave-function renormalization function,~$Z(p^2)$:
\begin{equation}S(p)=\frac{{\rm i}\,Z(p^2)}{\not\!p-M(p^2)+{\rm
i}\,\varepsilon}\ ,\qquad\not\!p\equiv p^\mu\,\gamma_\mu\
,\qquad\varepsilon\downarrow0\ .\label{Eq:FP}\end{equation}From
this relationship, we may conclude \cite{WL15} that, in
\emph{Euclidean-space\/} formulation, indicated by underlined
coordinates, the \emph{Bethe--Salpeter amplitudes\/} of massless
pseudoscalar mesons in the center-of-momentum frame
($\underline{\bm{P}}=0$) are controlled by $M(\underline{k}^2)$
and $Z(\underline{k}^2)$ according~to$$\Phi(\underline{k},0)
\propto\frac{Z(\underline{k}^2)\,M(\underline{k}^2)}
{\underline{k}^2+M^2(\underline{k}^2)}\,\underline{\gamma}_5
+\mbox{subleading contributions}\ .$$Ignoring the comparatively
minor variation of $Z(\underline{k}^2)$ with $\underline{k}^2$
leads to the sought relation \cite{WL15}
\begin{equation}\displaystyle\Phi(\underline{k},0)\propto
\frac{M(\underline{k}^2)}{\underline{k}^2+M^2(\underline{k}^2)}\,
\underline{\gamma}_5+\cdots\ .\label{Eq:SP}\end{equation}

In Ref.~\cite{WL15}, we used as our first piece of information
about $M(\underline{k}^2)$ a kind of by-product of a
QCD-compatible model calculation \cite{PM97b} of the quark
propagator $S(\underline{k})$ in Euclidean space, viz.\ that
\emph{in the chiral limit\/} the quark mass function decays, for
large $\underline{k}^2,$ essentially~like~$\underline{k}^{-2}$:
$$\lim_{\underline{k}^2\to\infty}M(\underline{k}^2)\propto
\frac{1}{\underline{k}^2}\qquad\Longrightarrow\qquad
\lim_{\underline{k}^2\to\infty}\Phi(\underline{k},0)\propto
\lim_{\underline{k}^2\to\infty}\frac{M(\underline{k}^2)}
{\underline{k}^2}\,\underline{\gamma}_5\propto
\frac{1}{\underline{k}^4}\,\underline{\gamma}_5\ .$$Introducing a
parameter $\mu$ with the dimension of mass, and guided by our
preference~for an analytic treatment if feasible, we modelled the
above feature of $\Phi(\underline{k},0)$ by the
simple~ansatz\begin{equation}\Phi(\underline{k},0)=\frac{1}
{(\underline{k}^2 +\mu^2)^2}\,\underline{\gamma}_5\ .\label{Eq:UV}
\end{equation}Integration w.r.t.\ the Euclidean-time
$\underline{k}$ component gave, for the Salpeter
component~$\varphi_2(p),$$$
\varphi_2(p)=4\,\sqrt{\frac{\mu^3}{\pi}}\,\frac{1}{(p^2+\mu^2)^{3/2}}\
,\qquad\mu>0\ ,\qquad\|\varphi_2\|^2\equiv\int\limits_0^\infty{\rm
d}p\,p^2\,|\varphi_2(p)|^2=1\ ,$$which, in \emph{configuration
space\/}, is just the modified Bessel function of second kind
$K_0(z)$ \cite{AS}:$$\varphi(r)=\frac{4\,\sqrt{2\,\mu^3}}
{\pi}\,K_0(\mu\,r)\ ,\qquad\mu>0\ ,\qquad\|\varphi\|^2\equiv
\int\limits_0^\infty{\rm d}r\,r^2\,|\varphi(r)|^2=1\ .$$For the
kinetic term $T(r)$ and the potential $V(r),$ this ansatz yields
\cite{WL15}, e.g., for $\mu=m>0,$
$$T(r)=\frac{2\,\sqrt{2\,m^3}}{r}\exp(-m\,r)\ ,\qquad
V(r)=-\frac{\pi}{2}\,\frac{\exp(-m\,r)}{r\,K_0(m\,r)}\ .$$

In the following, as our evident next step we improve the somewhat
na\"ive ansatz~(\ref{Eq:UV}) by accommodating, in addition to the
quark mass function's asymptotic behaviour considered above, the
needs of colour confinement as disclosed by axiomatic quantum
field theory \cite{QP}.

\section{Consequences of the Axiom of Reflection Positivity}
\label{Sec:ARP}Having taken advantage from our knowledge of the
ultraviolet behaviour of the quark mass function, logically our
next move must be to fathom the implications of colour confinement
for any potential $V(r)$ specifying an interaction kernel
$K(\bm{p},\bm{q})$ of the Salpeter~equation~(\ref{Eq:SE}).

\subsection{Confinement and analytic properties of Schwinger
functions}\label{sec:SF}From the point of view of experiment,
colour confinement forms a just heuristic description of the
empirically established fact of non-observation of isolated
coloured particles, i.e., the absence of the fundamental coloured
degrees of freedom of quantum chromodynamics from the spectrum of
observable states. Within the framework of quantum field theory, a
precise definition of colour confinement may be formulated
\cite{CDR92,CDR08} in terms of Schwinger functions, specific (not
time-ordered) analytic $n$-point functions of field operators in
Euclidean space, by analytic continuation of the Wightman
functions of \emph{axiomatic quantum field theory\/} \cite{QP}. By
the Osterwalder--Schrader reconstruction theorem \cite{OS}, each
Schwinger function related to an element of the Hilbert space of
observables fulfills the axiom of reflection positivity. In other
words, compliance with the axiom of reflection positivity forms a
necessary~condition for such relationship of Schwinger functions
to elements of the Hilbert space of observables.

For a \emph{two-point Schwinger function}, the fulfilment of the
axiom of reflection positivity is equivalent to the existence of a
K\"all\'en--Lehmann representation of this Schwinger function.
This, in turn, forbids inflexion points of this Schwinger function
at spacelike momenta \cite{CDR08}. Phrased the other way round, a
two-point Schwinger function with any such inflexion~point
violates the axiom of reflection positivity and therefore cannot
correspond to an element in the Hilbert space of observables: the
respective degree of freedom is subject to confinement.

Our aim is to confine quarks, with mass function
$M(\underline{k}^2),$ inside a Goldstone-type meson with
bound-state mass $\widehat{M}=0,$ that is, to prevent the quarks
from entering the spectrum of observables by hindering them to
propagate to infinity. In other words, we must assure that a quark
propagator is not the propagator of an observable particle.
According to the above, one circumstance which guarantees this is
the violation of the axiom of reflection positivity by the
(Euclidean-space) quark two-point Schwinger function, connected
to~the propagator associated to the quark in Minkowski space. Now,
the K\"all\'en--Lehmann representation of a Schwinger function
severely constrains its momentum-space behaviour in a variety of
ways. Among others, it does not tolerate the presence of an
inflexion point at spacelike momenta. Hence, a possibility to
achieve such confinement-enforcing breach of the axiom of
reflection positivity is to assume for the two-point Schwinger
function a behaviour incompatible~with at least one of the simple
constraints imposed on the analytical properties of any two-point
Schwinger function by the sheer fact of its \emph{possession\/} of
a K\"all\'en--Lehmann representation. Of course, confinement may
originate from other roots or reveal itself in a different manner.

For the purpose of the present investigation, we would like to
place our full wager~on the occurrence of an inflexion point.
Following an admittedly rather heuristic line of argument, let us
assume that the confinement-promoting properties of the quark
two-point Schwinger function will be carried over to the quark
propagator (\ref{Eq:FP}) in Euclidean-space representation and let
us, tentatively, attribute the latter characteristics to some
appropriate behaviour of the quark mass function
$M(\underline{k}^2).$ Then, by Eq.~(\ref{Eq:SP}), resulting from
the relationship between (two-point) quark propagator and
(three-point) quark--meson vertex implied \cite{PM97a,PM97b} by
the axial-vector Ward--Takahashi identity of QCD, such facets will
eventually get imprinted on the Bethe--Salpeter amplitude
$\Phi(\underline{k},0)$ of the Goldstone-type quark--antiquark
bound~state.

\subsection{Implementation of some consequences of colour
confinement}\label{sec:IPA}In view of the fundamental
interrelationships, compendiously sketched in Sec.~\ref{sec:SF},
between the qualitative pointwise behaviour of propagators in
momentum-space representation and the occurrence of confinement of
the associated elementary excitations, we now consider an ansatz
for the Bethe--Salpeter amplitude compatible with the requirements
of confinement.

Clearly, by way of our starting point, Eq.~(\ref{Eq:SP}), ensuring
colour confinement by asserting the existence of an inflexion
point in the quark mass function $M(\underline{k}^2)$ will be
reflected by the pointwise behaviour of the associated
Bethe--Salpeter amplitude $\Phi(\underline{k},\underline{P}).$
Therefore, in the center-of-momentum frame of the bound state, all
these considerations, combined~with our bias towards analytic
manageability, suggest as ansatz for $\Phi(\underline{k},0)$ of
still rather simple~form\begin{equation}\Phi(\underline{k},0)
=\left[\frac{1}{(\underline{k}^2+\mu^2)^2}
+\frac{\eta\,\underline{k}^2}{(\underline{k}^2+\mu^2)^3}\right]
\underline{\gamma}_5\ ,\qquad\mu>0\ ,\qquad\eta\in\mathbb{R}\
.\label{Eq:IPA}\end{equation}An integration with respect to the
momentum component $\underline{k}_4$ yields the Salpeter
amplitude$$\phi(\bm{k})\equiv\frac{1}{2\pi}\int{\rm
d}\underline{k}_4\,\Phi(\underline{k})=
\frac{4\,(1+\eta)\,\bm{k}^2+(4+\eta)\,\mu^2}
{16\,(\bm{k}^2+\mu^2)^{5/2}}\,\underline{\gamma}_5
\propto\varphi_2(|\bm{k}|)\,\underline{\gamma}_5$$and thus,
according to Eq.~(\ref{Eq:PSA}), the normalized Salpeter function
$\varphi_2(p)$ we are interested~in:\begin{equation}\varphi_2(p)
=\sqrt{\frac{256\,\mu^3}{\pi\,[256+\eta\,(320+109\,\eta)]}}\,
\frac{4\,(1+\eta)\,p^2+(4+\eta)\,\mu^2}{(p^2+\mu^2)^{5/2}}\
,\qquad\|\varphi_2\|^2=1\ .\label{Eq:SCp}\end{equation}So, in
momentum space our Salpeter component exhibits a smooth behaviour.
It assumes~a (for all $\eta$ real) finite value at the origin
$p=0,$ and vanishes, of course, in the limit
of~large~$p$:$$\varphi_2(0)=\frac{16\,(4+\eta)}
{\sqrt{\pi\,[256+\eta\,(320+109\,\eta)]\,\mu^3}}\
,\qquad\varphi_2(p)\xrightarrow[p\to\infty]{}0\ .$$

In configuration-space representation, this Salpeter-component
ansatz is expressible by means of two modified Bessel functions
$K_n(z)$ of the second kind \cite{AS} of the orders~$n=0,1$:
\begin{equation}\varphi(r)=\frac{16}{\pi}\,
\sqrt{\frac{2\,\mu^3}{256+\eta\,(320+109\,\eta)}}\,
[4\,(1+\eta)\,K_0(\mu\,r)-\eta\,\mu\,r\,K_1(\mu\,r)]\
,\qquad\|\varphi\|^2=1\ .\label{Eq:SCr}\end{equation}The two
specific values $\eta=0$ and $\eta=-1$ of the mixing parameter
constitute critical points of the inversion formalism utilized
here \cite{WL13} in the sense that for these two exceptional
values no longer an interplay of the two contributions to the
right-hand side of Eq.~(\ref{Eq:SCr}) but rather the first or
second term alone determines the behaviour of the
configuration-space Salpeter component function $\varphi(r),$ and
therefore nature and shape of the interaction~potential $V(r).$
Accordingly, in all following claims these two critical values of
$\eta$ deserve separate~attention.

Obviously, $\varphi(r)$ diverges at the origin $r=0,$ except for
$\eta=-1,$ and vanishes for~large~$r$:\begin{align*}\varphi(0)&
=\frac{16}{3\pi}\,\sqrt{\frac{2\,\mu^3}{5}}&&\mbox{for\ }\eta=-1\
,\\\varphi(r)&\xrightarrow[r\to0]{}-\frac{64\,(1+\eta)}{\pi}\,
\sqrt{\frac{2\,\mu^3}{256+\eta\,(320+109\,\eta)}}\ln(\mu\,r)
\xrightarrow[r\to0]{}\infty&&\mbox{for\ }\eta\ne-1\ ;\\
\varphi(r)&\xrightarrow[r\to\infty]{}0\ .\end{align*}

Due to the exponential decay of the modified Bessel functions of
the second kind $K_n(z),$ $n=0,1,2,\dots,$ the configuration-space
Salpeter component $\varphi(r)$ given by
Eq.~(\ref{Eq:SCr})~has~one, and only one, zero, $r_0,$ for each
choice of the mixing parameter $\eta$ lying in one of the
intervals $-\infty<\eta\lneqq-1$ or $0\lneqq\eta<\infty.$
Table~\ref{Tab:0} lists the numerical value of this zero~(in~units
of $1/\mu$) for several choices of $\eta$ considered in the
following. Barring the rather unlikely~possibility of an
accidental zero of the kinetic term $T(r)$ at the same location
$r=r_0,$ the division~by $\varphi(r)$ required by
Eq.~(\ref{Eq:Po}) forces the zero $r_0$ to induce a singularity of
the potential $V(r)$~at~$r=r_0.$

\begin{table}[hbt]\caption{Zero $r_0$ of the Salpeter component
(\ref{Eq:SCr}) for $\mu=1$ and various mixing
parameters~$\eta.$}\vspace{-.155ex}
\label{Tab:0}\begin{center}\begin{tabular}{lr}\hline\hline\\[-1.5ex]
\multicolumn{1}{c}{$\eta$}&\multicolumn{1}{c}{$r_0$ [$1/\mu$]}
\\[1.3ex]\hline\\[-1.5ex]0.5&11.51002562\dots\\1&7.51478467\dots
\\1.5&6.18423766\dots\\2&5.51940196\dots\\[0.9ex]\hline\hline
\end{tabular}$\qquad$\begin{tabular}{lr}\hline\hline\\[-1.5ex]
\multicolumn{1}{c}{$\eta$}&\multicolumn{1}{c}{$r_0$ [$1/\mu$]}
\\[1.3ex]\hline\\[-1.5ex]$-1.25$&0.41415074\dots\\$-1.5$&0.90800844\dots
\\$-1.75$&1.27449577\dots\\$-2$&1.55265125\dots\\[0.9ex]\hline\hline
\end{tabular}\end{center}\end{table}

\begin{figure}[hbt]\begin{center}\begin{tabular}{cc}
\psfig{figure=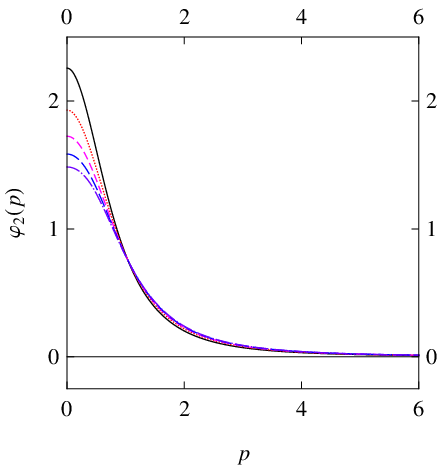,scale=1.7068}&
\psfig{figure=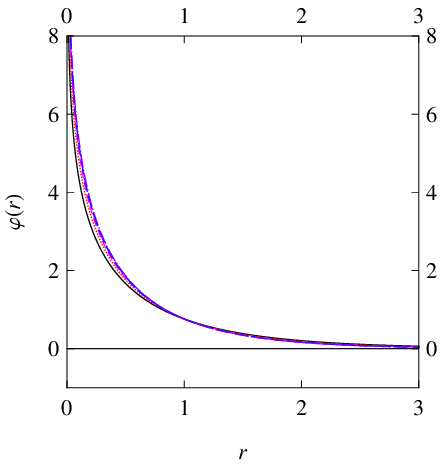,scale=1.7068}\\(a)&(b)\end{tabular}
\caption{Independent Salpeter component fully defining, at least
for any Salpeter equation (\ref{Eq:SE}) characterized by the
Lorentz structure $2\,\Gamma\otimes\Gamma=
\gamma_\mu\otimes\gamma^\mu+\gamma_5\otimes\gamma_5-1\otimes1$ of
the interaction kernel, the Salpeter amplitude (\ref{Eq:PSA}) for
pseudoscalar mesons, shown in adequate units of $\mu$ in its (a)
\emph{momentum-space\/} representation, $\varphi_2(p)\propto
(p^2+1)^{-3/2}+\eta\,(p^2+\frac{1}{4})\,(p^2+1)^{-5/2},$ and (b)
\emph{configuration-space\/} representation, $\varphi(r)\propto
K_0(r)+\eta\,[K_0(r)-r\,K_1(r)/4],$ at the value $\eta=0$ (black
solid line \cite[Sec.~4.C]{WL15}), $\eta=0.5$ (red dotted line),
$\eta=1$ (magenta short-dashed line), $\eta=1.5$ (blue long-dashed
line) and $\eta=2$ (violet dot-dashed
line)~of~our~mixing~parameter.}\label{Fig:SC}\end{center}
\end{figure}

Figure~\ref{Fig:SC} shows, in appropriate units of $\mu$ (or,
equivalently, for $\mu=1$), for a few~values of the mixing
parameter $\eta,$ the behaviour of our \emph{inflexion-friendly\/}
ansatz for the independent Salpeter component, in both momentum
space, Eq.~(\ref{Eq:SCp}), and configuration space,
Eq.~(\ref{Eq:SCr}).

The only remaining challenge is to deduce the kinetic term $T(r).$
In the following, this is done for two (trivially) analytically
accessible cases: $m=0$ (Sec.~\ref{sec:m=0}) and
$m=\mu$~(Sec.~\ref{sec:m=mu}).

\section{Analytic Configuration-Space Confining Potentials}
\label{Sec:Po}Fed with the ansatz (\ref{Eq:SCp}), our inversion
machinery promptly returns all potentials aimed~at.

In view of our discussion of the presence of one zero $r_0$ in the
ansatz (\ref{Eq:SCr}) for the~Salpeter component $\varphi(r)$ in
Sec.~\ref{sec:IPA}, we have to expect that, for any value of the
mixing parameter $\eta$ outside the ``safe'' interval
$-1\le\eta\le0,$ any potential $V(r)$ inferred in this way will
develop a singularity at the location of this zero $r_0.$ As a
matter of fact, this singularity will prove to be an infinite
discontinuity of $V(r)$ and $V(r)$ to exhibit one of the
following~two~behaviours:
\begin{align*}&V(r)\xrightarrow[r\to r_0^-]{}+\infty\ ,\qquad
V'(r)\xrightarrow[r\to r_0^-]{}+\infty\ ,\qquad
V(r)\xrightarrow[r\to r_0^+]{}-\infty \ ,\qquad
V'(r)\xrightarrow[r\to r_0^+]{}+\infty\ ;\\&V(r)\xrightarrow[r\to
r_0^-]{}-\infty\ ,\qquad V'(r)\xrightarrow[r\to r_0^-]{}-\infty\
,\qquad V(r)\xrightarrow[r\to r_0^+]{}+\infty \ ,\qquad
V'(r)\xrightarrow[r\to r_0^+]{}-\infty\ .\end{align*}The mixing
$\widetilde\eta$ where $V(r)$ flips from one to the other is fixed
by its competing contributions.

As a consequence, for $\eta\notin[-1,0],$ when tracing this
potential $V(r),$ for $r$ rising from the origin $r=0,$ one
encounters an infinite discontinuity situated at $r=r_0,$ i.e., a
region where this potential $V(r)$ grows beyond bounds.
Accordingly, for interquark separations $r$ smaller than the
$\varphi(r)$ zero $r_0,$ that is, in the region $0\le r<r_0,$ the
potential $V(r)$~can be claimed to exhibit a confining behaviour.
In this spirit, the location $r_0$ of the singularity of $V(r)$
off the spatial origin $r=0$ might be interpreted as introducing
some sort of~``confinement~radius.''

\subsection{Analytic treatment: bound-state constituents of mass
$\bm{m=0}$}\label{sec:m=0}For massless quarks, i.e., if $m=0,$ the
configuration-space kinetic term $T(r)$ corresponding to the
Salpeter component (\ref{Eq:SCr}) can be expressed in terms of
both modified Bessel functions of the first kind \cite{AS}
$I_n(z)$ for $n=0,1$ and modified Struve functions \cite{AS} ${\bf
L}_n(z)$ for $n=0,1$:\begin{align*}T(r)&
=\sqrt{\frac{2\,\mu^3}{256+\eta\,(320+109\,\eta)}}\,\frac{8}{\pi\,r}
\{\pi\,[4+\eta\,(4+\mu^2\,r^2)]\,[I_0(\mu\,r)-{\bf L}_0(\mu\,r)]\\
&+\pi\,(4+5\,\eta)\,\mu\,r\,[I_1(\mu\,r)-{\bf L}_1(\mu\,r)]
-4\,(2+3\,\eta)\,\mu\,r\}\ .\end{align*}In the extracted
interaction potential, the nasty normalization factor necessarily
drops~out:
\begin{align}V(r)=-&\{\pi\,[4+\eta\,(4+\mu^2\,r^2)]\,[I_0(\mu\,r)
-{\bf L}_0(\mu\,r)]+\pi\,(4+5\,\eta)\,\mu\,r\,[I_1(\mu\,r)-{\bf
L}_1(\mu\,r)]\nonumber\\&-4\,(2+3\,\eta)\,\mu\,r\}\,
/\,\{2\,r\,[4\,(1+\eta)\,K_0(\mu\,r)-\eta\,\mu\,r\,K_1(\mu\,r)]\}\
.\label{Eq:Po0}\end{align}With the relation $\pi\,{\bf
L}_1(z)+2=\pi\,{\bf L}_{-1}(z)$ [e.g., Eq.~(12.2.4) of
Ref.~\cite{AS} for $\nu=0$],~it is trivial to show that, for
$\eta=0,$ this potential $V(r)$ reduces to that one found in
Sec.~V.A~of~Ref.~\cite{WL15}. At the origin $r=0,$ this potential
$V(r)$ develops, except for $\eta=-1,$ a Coulomb singularity
governed by the first term in its denominator and logarithmically
softened by the divergent behaviour, $K_0(x\to0)\approx-\ln(x),$
of the modified Bessel function of the second kind $K_0(x)$:
\begin{align*}V(0)&=-2\,\mu&&\mbox{for\ }\eta=-1\ ,\\
V(r)&\xrightarrow[r\to0]{}\frac{\pi}{2\,r\ln(\mu\,r)}
\xrightarrow[r\to0]{}-\infty&&\mbox{for\ }\eta\ne-1\ .\end{align*}
With the exception of the case $\eta=-1,$ for which this potential
$V(r)$ assumes a finite~value, the independence of this
short-distance behaviour of $V(r)$ from the parameter $\eta$
controlling the amount of admixture of the second term in our
ansatz (\ref{Eq:IPA}) renders clear the irrelevance of the second
term in the denominator of $V(r)$ for the behaviour of $V(r)$ at
the origin~$r=0.$

\begin{figure}[hbt]\begin{center}\psfig{figure=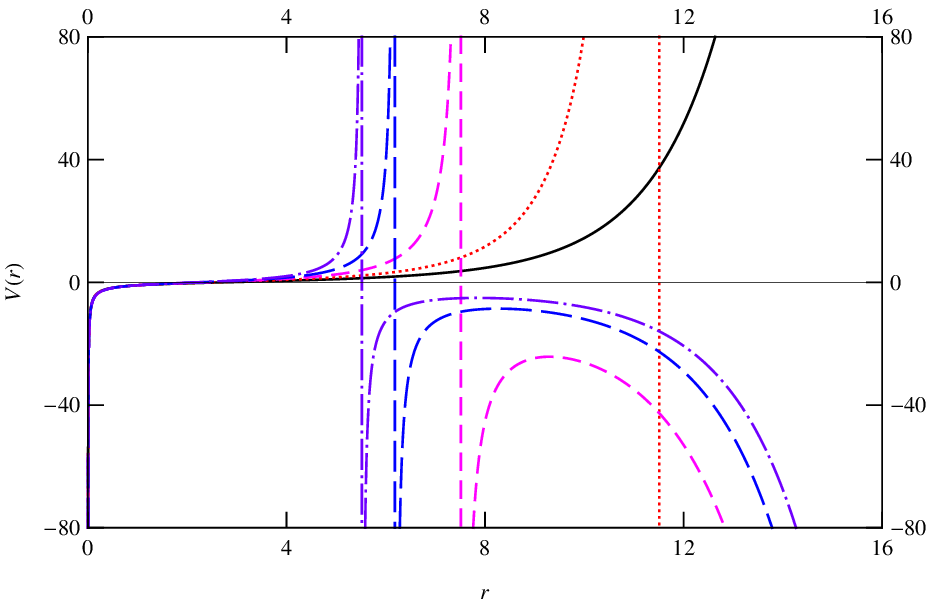,scale=1.7}
\caption{Configuration-space potential $V(r)$ extracted from the
Salpeter equation (\ref{Eq:SE}) with interaction-kernel Lorentz
structure $2\,\Gamma\otimes\Gamma=\gamma_\mu\otimes\gamma^\mu
+\gamma_5\otimes\gamma_5-1\otimes1$ by assuming the ansatz
$\varphi_2(p)\propto(p^2+1)^{-3/2}+\eta\,(p^2+\frac{1}{4})\,
(p^2+1)^{-5/2}$ for the nonvanishing component of the Salpeter
amplitude (\ref{Eq:PSA}) to describe massless pseudoscalar bound
states of fermions with mass $m=0$: $V(r)=-N(r)/D(r)$ with the two
abbreviations
$D(r)\equiv2\,r\,[4\,(1+\eta)\,K_0(r)-\eta\,r\,K_1(r)]$ and
$N(r)\equiv\pi\,[4+\eta\,(4+r^2)]\,[I_0(r)-{\bf L}_0(r)]
+\pi\,(4+5\,\eta)\,r\,[I_1(r)-{\bf L}_1(r)]-4\,(2+3\,\eta)\,r$ for
notational ease introduced for denominator and numerator,
respectively, depicted for the values $\eta=0$ (black solid line
\cite[Sec.~5.A]{WL15}), $\eta=0.5$ (red dotted line), $\eta=1$
(magenta short-dashed line), $\eta=1.5$ (blue long-dashed line)
and $\eta=2$ (violet dot-dashed line)~of~our~mixing~parameter.}
\label{Fig:Vm=0}\end{center}\end{figure}

\begin{figure}[hbt]\begin{center}
\psfig{figure=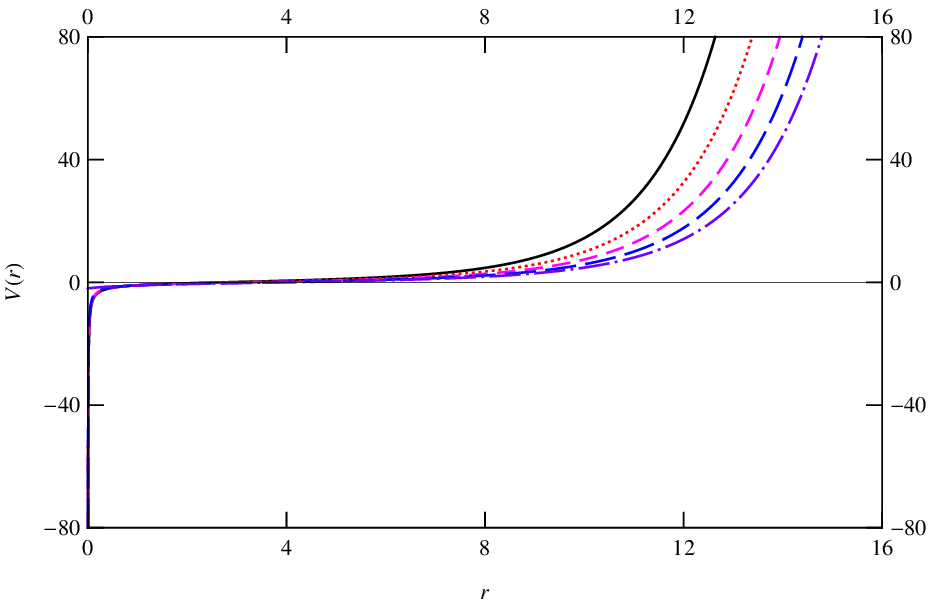,scale=1.7}
\caption{Configuration-space potential $V(r)$ extracted from the
Salpeter equation (\ref{Eq:SE}) with interaction-kernel Lorentz
structure $2\,\Gamma\otimes\Gamma=\gamma_\mu\otimes\gamma^\mu
+\gamma_5\otimes\gamma_5-1\otimes1$ by assuming the ansatz
$\varphi_2(p)\propto(p^2+1)^{-3/2}+\eta\,(p^2+\frac{1}{4})\,
(p^2+1)^{-5/2}$ for the nonvanishing component of the Salpeter
amplitude (\ref{Eq:PSA}) to describe massless pseudoscalar bound
states of fermions with mass $m=0,$ $V(r)=-N(r)/D(r)$ with the two
abbreviations
$D(r)\equiv2\,r\,[4\,(1+\eta)\,K_0(r)-\eta\,r\,K_1(r)]$ and
$N(r)\equiv\pi\,[4+\eta\,(4+r^2)]\,[I_0(r)-{\bf L}_0(r)]
+\pi\,(4+5\,\eta)\,r\,[I_1(r)-{\bf
L}_1(r)]-4\,(2+3\,\eta)\,r$~for~numerator and denominator,
respectively, for a few $\eta$-parameter choices from the interval
$-1\le\eta\le0$: $\eta=0$ (black solid line, again
\cite[Sec.~5.A]{WL15}), $\eta=-0.25$ (red dotted line),
$\eta=-0.5$ (magenta short-dashed line), $\eta=-0.75$ (blue
long-dashed line), and $\eta=-1$ (violet dot-dashed line).}
\label{Fig:Vm=0nz}\end{center}\end{figure}

\begin{figure}[hbt]\begin{center}
\psfig{figure=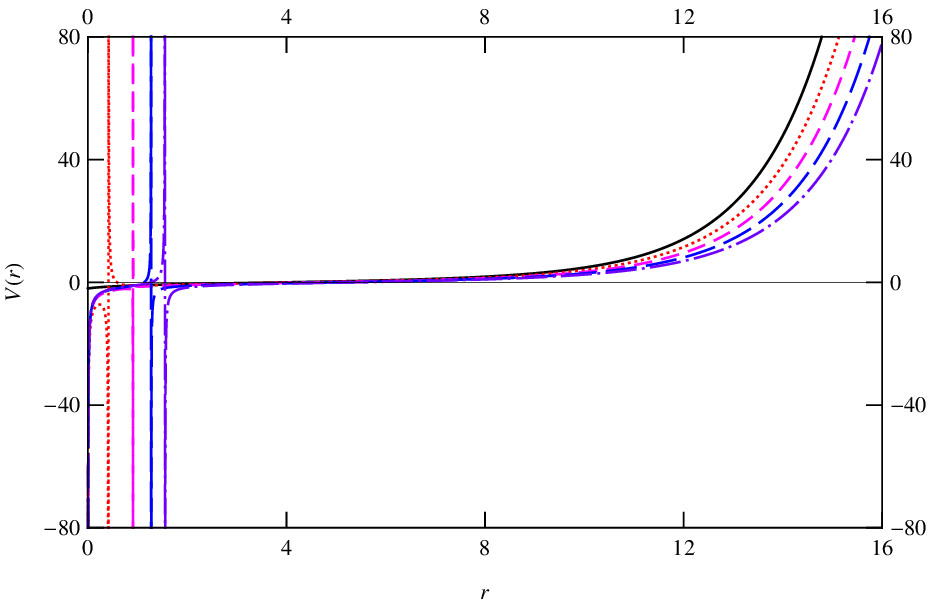,scale=1.7}
\caption{Configuration-space potential $V(r)$ extracted from the
Salpeter equation (\ref{Eq:SE}) with interaction-kernel Lorentz
structure $2\,\Gamma\otimes\Gamma=\gamma_\mu\otimes\gamma^\mu
+\gamma_5\otimes\gamma_5-1\otimes1$ by assuming the ansatz
$\varphi_2(p)\propto(p^2+1)^{-3/2}+\eta\,(p^2+\frac{1}{4})\,
(p^2+1)^{-5/2}$ for the nonvanishing component of the Salpeter
amplitude (\ref{Eq:PSA}) to describe massless pseudoscalar bound
states of fermions with mass $m=0,$ $V(r)=-N(r)/D(r)$ with the two
abbreviations
$D(r)\equiv2\,r\,[4\,(1+\eta)\,K_0(r)-\eta\,r\,K_1(r)]$ and
$N(r)\equiv\pi\,[4+\eta\,(4+r^2)]\,[I_0(r)-{\bf L}_0(r)]
+\pi\,(4+5\,\eta)\,r\,[I_1(r)-{\bf
L}_1(r)]-4\,(2+3\,\eta)\,r$~for~numerator and denominator,
respectively, for a few $\eta$-parameter values from the range
$-\infty<\eta\le-1$: $\eta=-1$ (black solid line, as a benchmark),
$\eta=-1.25$ (red dotted line), $\eta=-1.5$ (magenta short-dashed
line), $\eta=-1.75$ (blue long-dashed line), and $\eta=-2$ (violet
dot-dashed line).} \label{Fig:Vm=0ne}\end{center}\end{figure}

For the distance $r$ rising from zero to $\infty,$ the qualitative
behaviour of the potential (\ref{Eq:Po0}), exemplified in
Figs.~\ref{Fig:Vm=0} through \ref{Fig:Vm=0ne} in units of
appropriate powers of $\mu$ (which is tantamount~to setting
$\mu=1$), varies among regions of $\eta$ separated by the critical
points $\eta=0$ and $\eta=-1$:\begin{itemize}\item For $\eta>0$
(Fig.~\ref{Fig:Vm=0}), $V(r)$ rises for increasing $r$
monotonically to $\infty$ at the discontinuity at $r=r_0,$ where
it jumps from $+\infty$ to $-\infty$ and then remains below zero
up to $r=\infty.$\item For $-1<\eta\le0$ (Fig.~\ref{Fig:Vm=0nz}),
in spite of the absence of zeros in $\varphi(r)$ in this
case,~$V(r)$~rises for increasing $r$ monotonically from its
notorious singularity at $r=0$ to $\infty$ for large~$r.$\item For
$\eta\le-1$ (Fig.~\ref{Fig:Vm=0ne}), $V(r)$ is affected by its
``switching'' value of $\eta,$
$\widetilde\eta\approx-1.520216412$:\begin{itemize}\item For
$\widetilde\eta<\eta<-1,$ with rising $r$ $V(r)$ first stays below
zero up to $r_0,$ there~it~jumps from $-\infty$ to $+\infty,$ then
passes a local minimum and finally rises to $\infty$
for~large~$r.$\item For $-\infty<\eta<\widetilde\eta,$ for $r$
approaching the $\varphi(r)$-induced discontinuity from~the left
$V(r)$ rises to $\infty,$ then jumps from $+\infty$ to $-\infty,$
and rises again, for $r\to\infty,$~to~$\infty.$\end{itemize}
\end{itemize}With due satisfaction we find that, for massless
quarks, the ansatz (\ref{Eq:IPA}) defines, irrespective of the
choice of the mixing $\eta,$ for one reason or the other a
potential capable of confinement.

\subsection{Analytic result: bound-state constituents of mass
$\bm{m=\mu>0}$}\label{sec:m=mu}In case the bound-state
constituents' mass $m$ equals precisely the mass $\mu$ that
parametrizes our ansatz (\ref{Eq:IPA}), that is, for $\mu=m>0,$
the numerator and denominator of the integrand in the
Fourier--Bessel transform of the kinetic term $E(p)\,\varphi_2(p)$
resemble each other to a degree that enables $T(r)$ to be an
analytic expression with Yukawa and exponential~contributions:
$$T(r)=\frac{4\,\sqrt{2\,m^3}\,[8+\eta\,(8-3\,m\,r)]}
{\sqrt{256+\eta\,(320+109\,\eta)}}\,\frac{\exp(-m\,r)}{r}\
.$$Accordingly, our prototype potential for the case of massive
bound-state constituents~reads
\begin{align}V(r)&=-\frac{\pi\,[8+\eta\,(8-3\,m\,r)]\exp(-m\,r)}
{4\,r\,[4\,(1+\eta)\,K_0(m\,r)-\eta\,m\,r\,K_1(m\,r)]}\
.\label{Eq:Po1}\end{align}Similarly to the case $m=0$ studied in
Sec.~\ref{sec:m=0}, and for the same reasons, except for~$\eta=-1$
the potential $V(r)$ has its logarithmically softened Coulomb
singularity at the origin $r=0$:\begin{align*}
V(0)&=-3\pi\,m/4=-m\times2.35619449\ldots&&\mbox{for\ }\eta=-1\
,\\V(r)&\xrightarrow[r\to0]{}\frac{\pi}{2\,r\ln(m\,r)}
\xrightarrow[r\to0]{}-\infty&&\mbox{for\ }\eta\ne-1\ .\end{align*}
In the limit $r\to\infty,$ all $V(r)$ approach zero, with decay
controlled for $\eta=0$ by the~modified Bessel function $K_0$
\cite{WL15} but for $\eta\ne0$ by the modified Bessel function
$K_1$ in the denominator:
\begin{align*}V(r)&=-\frac{\pi}{2}\,\frac{\exp(-m\,r)}{r\,K_0(m\,r)}
\xrightarrow[r\to\infty]{}-\sqrt{\frac{\pi\,m}{2\,r}}
\xrightarrow[r\to\infty]{}0&&\mbox{for\ }\eta=0\mbox{\cite{WL15}}\
,\\V(r)&\xrightarrow[r\to\infty]{}
-\frac{3\pi}{4}\,\frac{\exp(-m\,r)}{r\,K_1(m\,r)}
\xrightarrow[r\to\infty]{}-\frac{3}{2}\,\sqrt{\frac{\pi\,m}{2\,r}}
\xrightarrow[r\to\infty]{}0&&\mbox{for\ }\eta\ne0\ .\end{align*}

\begin{figure}[hbt]\begin{center}\psfig{figure=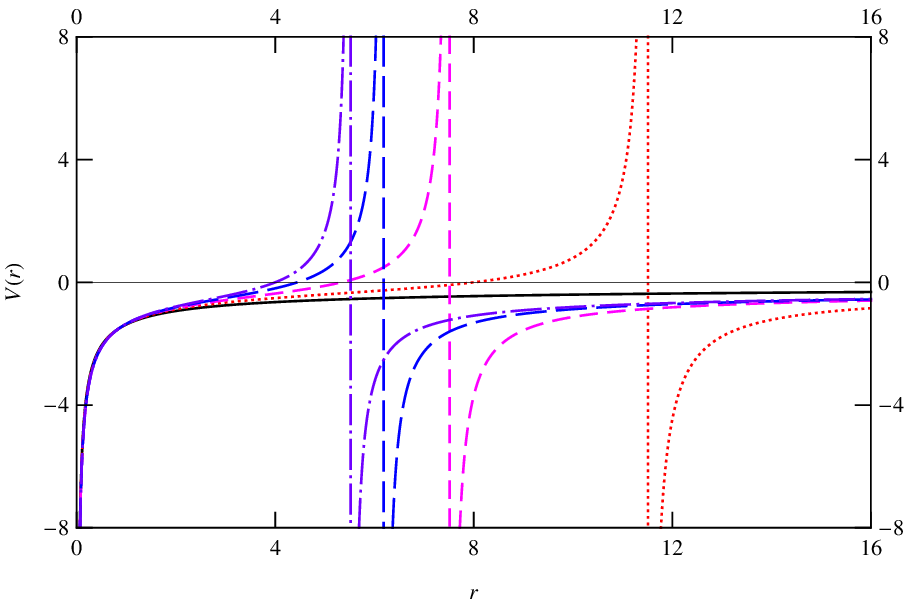,scale=1.7}
\caption{Configuration-space potential $V(r)$ extracted from the
Salpeter equation (\ref{Eq:SE}) with interaction-kernel Lorentz
structure $2\,\Gamma\otimes\Gamma=\gamma_\mu\otimes\gamma^\mu
+\gamma_5\otimes\gamma_5-1\otimes1$ by assuming~the~ansatz
$\varphi_2(p)\propto(p^2+1)^{-3/2}+\eta\,(p^2+\frac{1}{4})\,
(p^2+1)^{-5/2}$ for the nonvanishing component of the~Salpeter
amplitude (\ref{Eq:PSA}) to describe massless pseudoscalar bound
states of fermions with mass $m=1$:
$V(r)=-\{\pi\,[8+\eta\,(8-3\,r)]\exp(-r)\}\,/\,
\{4\,r\,[4\,(1+\eta)\,K_0(r)-\eta\,r\,K_1(r)]\},$ for the
values~$\eta=0$ (black solid line \cite[Sec.~5.B]{WL15}),
$\eta=0.5$ (red dotted line), $\eta=1$ (magenta short-dashed
line), $\eta=1.5$ (blue long-dashed line) and $\eta=2$ (violet
dot-dashed line) of our mixing~parameter.}\label{Fig:Vm=mu}
\end{center}\end{figure}

\begin{figure}[hbt]\begin{center}\psfig{figure=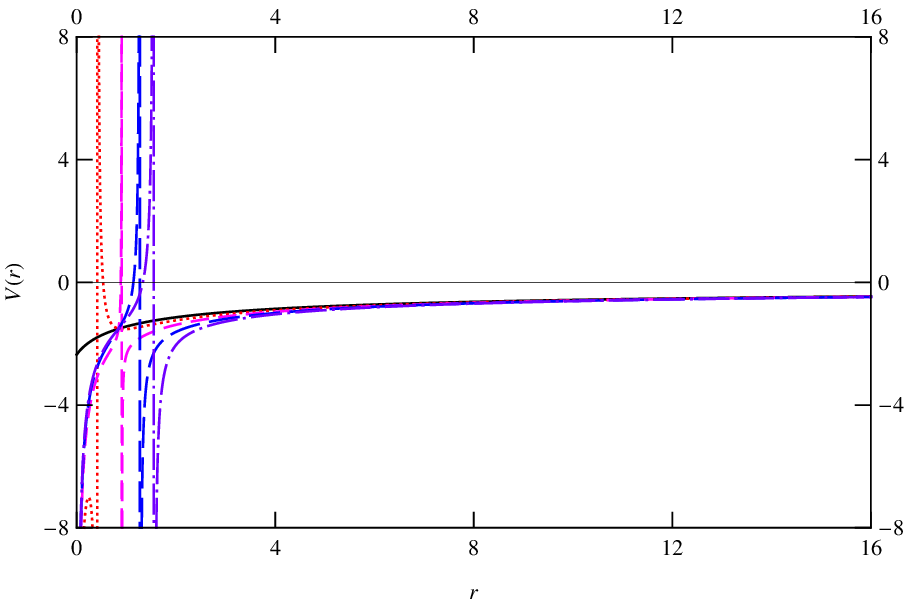,scale=1.7}
\caption{Configuration-space potential $V(r)$ extracted from the
Salpeter equation (\ref{Eq:SE}) with interaction-kernel Lorentz
structure $2\,\Gamma\otimes\Gamma=\gamma_\mu\otimes\gamma^\mu
+\gamma_5\otimes\gamma_5-1\otimes1$ by assuming the ansatz
$\varphi_2(p)\propto(p^2+1)^{-3/2}+\eta\,(p^2+\frac{1}{4})\,
(p^2+1)^{-5/2}$ for the nonvanishing component of the Salpeter
amplitude (\ref{Eq:PSA}) to describe massless pseudoscalar bound
states of fermions with mass $m=1,$
$V(r)=-\{\pi\,[8+\eta\,(8-3\,r)]\exp(-r)\}\,/\,
\{4\,r\,[4\,(1+\eta)\,K_0(r)-\eta\,r\,K_1(r)]\},$~for~``confining,''
yet negative mixing: $\eta=-1$ (black solid line), $\eta=-1.25$
(red dotted line), $\eta=-1.5$~(magenta short-dashed line),
$\eta=-1.75$ (blue long-dashed line), and $\eta=-2$ (violet
dot-dashed line).}\label{Fig:Vm=mu-ne}\end{center}\end{figure}

The potential (\ref{Eq:Po1}) rises to $\infty$ in precisely those
regions of $\eta$ that enforce a zero of $\varphi(r),$ as
depicted, for $\mu=1,$ in Figs.~\ref{Fig:Vm=mu} and
\ref{Fig:Vm=mu-ne} for those cases where a \emph{confining\/}
behaviour~is~found:\begin{itemize}\item For $\eta>0$
(Fig.~\ref{Fig:Vm=mu}), $V(r)$ fulfils all confining obligations
by growing beyond bounds for $r\nearrow r_0,$ then drops from
$+\infty$ to $-\infty,$ and finally approaches zero in the
limit~$r\to\infty.$\item For $-1\le\eta\le0,$ in which case $V(r)$
does not encounter a $\varphi(r)$-related singularity, the
monotonic rise with $r$ to the asymptotic value $V(\infty)=0$
betrays lack of confinement.\item For $\eta<-1$
(Fig.~\ref{Fig:Vm=mu-ne}), the jump of $V(r)$ at its discontinuity
flips for $\widetilde\eta\approx-1.463012572$:\begin{itemize}\item
For $\widetilde\eta<\eta<-1,$ $V(r)$ remains strictly negative up
to $r_0,$ performs a $-\infty$~to~$+\infty$ jump and approaches in
the limit $r\to\infty,$ via a local minimum, zero from~below.\item
For $-\infty<\eta<\widetilde\eta,$ $V(r)$ rises with $r$ at its
inevitable jump discontinuity at~$r=r_0$ to $+\infty,$ succumbs to
a fall to $-\infty$ and resumes its rise to zero in the far
distance.\end{itemize}\end{itemize}We are led to conclude that a
proper incorporation of confinement, manifesting itself by the
presence of an infinite discontinuity at the location of the zero
$r_0$ of the configuration-space Salpeter amplitude $\varphi(r),$
entails a drastic alteration of the physical impact of
the~extracted potential $V(r)$: whereas in Ref.~\cite{WL15},
starting from a Salpeter-amplitude ansatz equivalent to the
special case of a vanishing admixture of the second term in our
present ansatz~(\ref{Eq:IPA}),~i.e., for $\eta=0,$ we observed
that not only for $m=\mu>0$ but, in fact, for all $m\ge\mu$
the~extracted potential is not confining, here we obtain also for
quarks with nonzero mass, at least for the case $m=\mu,$
confinement for any mixing $\eta\notin[-1,0]$ enabling the
existence of a~zero~of~$\varphi(r).$

\section{Summary and Discussion of Findings, and Outlook}
\label{Sec:SCO}In the present study, we demonstrated, within an
instantaneous Bethe--Salpeter formalism, that it is achievable to
formulate, for pseudoscalar mesons, exact analytical solutions of
the homogeneous Bethe--Salpeter equation for quark--antiquark
bound states that exhibit both the absolute confinement demanded
from coloured degrees of freedom and the masslessness expected for
all Goldstone bosons related to spontaneously broken continuous
symmetries, in the sense of establishing a rigorous relationship
between tentatively postulated solutions and the form of the
effective interaction responsible for the formation of such bound
states.

By inversion, we harvest the effective interactions in the
disguise of configuration-space potentials that carry an imprint
of confinement:\footnote{One might wonder whether such a potential
provides absolute confinement. However: Due to the rise of the
potential to infinity in suitable intervals adjacent to the
off-origin discontinuity, we may~take as granted that any
transmission coefficient for the corresponding Schr\"odinger
problem vanishes. Since~the relativistic kinetic energy is bounded
from above by its nonrelativistic counterpart, entering the
Schr\"odinger equation, we expect that the transmission
coefficient of a spinless Salpeter equation with such potential
vanishes~too.} The absence of free coloured~states entails a zero
of the bound-state wave function, which causes the potentials to
rise without limit at finite interquark distance and thus appears
as \emph{the\/} crucial ingredient of the present~scenario.

Needless to say, our next step has to be to leave the
comparatively safe realm of~analytic investigations and to exploit
the explicit findings for the quark mass functions derived from
phenomenologically reliable QCD-based models within the framework
of Dyson--Schwinger equations existing in the literature
(unfortunately, however, available at present only as the results
of numerical computations) in order to obtain a still more
realistic understanding of the way the strong interactions enter
in the Salpeter approach to pions and kaons \cite{WL16ip}. Such
analysis might even culminate in an attempt to approximately
reconstruct the shape of the Bethe--Salpeter amplitudes for light
pseudoscalar mesons in form of an analytic~expression.

\section*{Acknowledgements}We are grateful to Lei Chang for
drawing our attention to an implication of the violation of the
axiom of reflection positivity for the behaviour of confining
Bethe--Salpeter amplitudes.


\small\begin{thebibliography}{30}
\bibitem{BSE}H.~A.~Bethe and E.~E.~Salpeter, Phys.~Rev.~{\bf
82} (1951) 309;\\M.~Gell-Mann and F.~Low, Phys.~Rev.~{\bf 84}
(1951) 350;\\E.~E.~Salpeter and H.~A.~Bethe, Phys.~Rev.~{\bf 84}
(1951) 1232.
\bibitem{WL05:IBSEWEP}W.~Lucha and F.~F.~Sch\"oberl, J.~Phys.~G:
Nucl.~Part.~Phys.~{\bf 31} (2005) 1133, arXiv:hep-th/0507281.
\bibitem{SE}E.~E.~Salpeter, Phys.~Rev.~{\bf 87} (1952) 328.
\bibitem{WL13}W.~Lucha and F.~F.~Sch\"oberl, Phys.~Rev.~D {\bf 87}
(2013) 016009, arXiv:1211.4716 [hep-ph].
\bibitem{WLPoS}W.~Lucha, Proc.~Sci., EPS-HEP 2013 (2013) 007,
arXiv:1308.3130 [hep-ph].
\bibitem{WL15}W.~Lucha and F.~F.~Sch\"oberl, Phys.~Rev.~D {\bf 92}
(2015) 076005, arXiv:1508.02951 [hep-ph].
\bibitem{WL07:HORSE}Z.-F.~Li, W.~Lucha, and F.~F.~Sch\"oberl,
Phys.~Rev.~D {\bf 76} (2007) 125028, arXiv:0707.3202 [hep-ph].
\bibitem{JFL}J.-F.~Laga\"e, Phys.~Rev.~D {\bf 45} (1992)
305.
\bibitem{OVW}M.~G.~Olsson, S.~Veseli, and K.~Williams, Phys.~Rev.~D
{\bf 52} (1995) 5141, arXiv:hep-ph/9503477.
\bibitem{AS}{\em Handbook of Mathematical Functions}, edited by
M.~Abramowitz and I.~A.~Stegun (Dover, New York, 1964).
\bibitem{WL-SSE}W.~Lucha and F.~F.~Sch\"oberl,
Int.~J.~Mod.~Phys.~A {\bf 07} (1992) 6431; in {\em Proceedings of
the International Conference on Quark Confinement and the Hadron
Spectrum\/}, edited by N.~Brambilla and G.~M.~Prosperi (World
Scientific, River Edge, NJ, 1995) p.~100, arXiv:hep-ph/9410221;
Int.~J.~Mod.~Phys.~A {\bf 14} (1999) 2309, arXiv:hep-ph/9812368;
Fizika B {\bf 8} (1999) 193, arXiv:hep-ph/9812526; Recent
Res.~Devel.~Phys.~{\bf 5} (2004) 1423, arXiv:hep-ph/0408184.
\bibitem{PM97a}P.~Maris, C.~D.~Roberts, and P.~C.~Tandy,
Phys.~Lett.~B {\bf 420} (1998) 267, arXiv:nucl-th/9707003.
\bibitem{PM97b}P.~Maris and C.~D.~Roberts, Phys.~Rev.~C {\bf 56}
(1997) 3369, arXiv:nucl-th/9708029.
\bibitem{QP}J.~Glimm and A.~Jaffe, \emph{Quantum Physics --- A
Functional Integral Point of View\/} (Springer, New York, 1981).
\bibitem{CDR92}C.~D.~Roberts, A.~G.~Williams, and G.~Krein,
Int.~J.~Mod.~Phys.~A {\bf 07} (1992) 5607.
\bibitem{CDR08}C.~D.~Roberts, Prog.~Part.~Nucl.~Phys.~{\bf 61}
(2008) 50, arXiv:0712.0633 [nucl-th].
\bibitem{OS}K.~Osterwalder and R.~Schrader,
Commun.~Math.~Phys.~{\bf 31} (1973) 83; {\bf 42} (1975) 281.
\bibitem{WL16ip}W.~Lucha and F.~F.~Sch\"oberl, in preparation.
\end{thebibliography}
\end{document}